\documentclass[aps,prb,twocolumn,superscriptaddress]{revtex4}
\usepackage{graphicx}
\usepackage{amsmath} 
\usepackage{amssymb} 
\usepackage{epsfig} 
 
\begin{document} 
 
\title{Giant vortices in rotating electron droplets}

\author{E.~R{\"a}s{\"a}nen} 
\email[Electronic address: ]{esa@physik.fu-berlin.de} 
\affiliation{Institut f{\"u}r Theoretische Physik,
Freie Universit{\"a}t Berlin, Arnimallee 14, D-14195 Berlin, Germany}
\affiliation{Laboratory of Physics, Helsinki University of Technology, 
P.O. Box 1100, FI-02015 HUT, Finland} 
\author{H.~Saarikoski} 
\affiliation{Kavli Institute of NanoScience, Delft University
of Technology, 2628 CJ Delft, the Netherlands}
\affiliation{Laboratory of Physics, Helsinki University of Technology, 
P.O. Box 1100, FI-02015 HUT, Finland} 
\author{Y.~Yu} 
\affiliation{Mathematical Physics, Lund Institute of Technology, 
SE-22100 Lund, Sweden} 
\author{A.~Harju} 
\affiliation{Laboratory of Physics, Helsinki University of Technology, 
P.O. Box 1100, FI-02015 HUT, Finland} 
\author{M.~J.~Puska} 
\affiliation{Laboratory of Physics, Helsinki University of Technology, 
P.O. Box 1100, FI-02015 HUT, Finland} 
\author{S.~M. Reimann} 
\affiliation{Mathematical Physics, Lund Institute of Technology, 
SE-22100 Lund, Sweden} 

\date{\today} 
 
\begin{abstract} 
We predict the formation of giant vortices in 
quasi-two-dimensional quantum dots at high magnetic 
fields, i.e., in rapidly rotating electron droplets.
Our numerical results of quantum dots confined
by a flat, anharmonic potential show ground states
where vortices are accumulated in the center of the 
dot, thereby leading to large cores in the electron 
and current densities. The phenomenon is analogous
to what was recently found in rotating Bose-Einstein 
condensates. The giant-vortex states leave measurable 
signatures in the ground-state energetics. 
The conditions for the giant-vortex formation as well as
the internal structure of the vortex cores are discussed.
\end{abstract}

\maketitle 

\section{Introduction}

In analogy to the well-known classical phenomenon, a vortex in a
quantum system can be characterized by rotational flow of particles,
forming a cavity at its center. It corresponds to a node in the wave
function, associated with a phase change of integer multiples of $2\pi $
for each path enclosing it. The formation of vortices in quantum liquids
has been observed, e.g., in superconductors,~\cite{super} He
liquids,~\cite{helium} and more recently, rotating Bose-Einstein
condensates (BECs) in atomic traps.~\cite{BEref}

Recently, electronic structure calculations of quasi-two-dimensional (2D)
quantum dots (QDs)~\cite{qd} have predicted formation of vortices in
rapidly rotating electron droplets.~\cite{henri1,tavernier,toreblad}
In this case an external magnetic field is used to set the system 
rotating. At sufficiently high fields the system tends to minimize
the total energy by nucleating vortices, i.e., the quanta of the magnetic 
flux. This highly correlated quantum system of electrons and hole-like vortex
quasi-particles~\cite{manninen} eventually shows formation of
composite particles comprising an electron and a number of 
vortices.~\cite{henri1} 
The fractional quantum Hall effect of
two-dimensional electron gas can be seen as bulk counterpart of this
phenomenon.~\cite{chakraborty} The physics of rotating fermions in quantum
dots shows remarkable similarities to that of rotating bosons in
BECs.~\cite{toreblad} This can be seen as a consequence of the
quantization of angular momentum in 2D systems.

High angular momentum of a rotating group of particles (bosons or
fermions) is associated with the formation of multiple vortices which 
can form as a lattice or a cluster inside the system. Another 
possibility is that
vortices are spatially concentrated into a small area at the center of
rotation. This structure is characterized by a large
localized core around which the phase shift is an integer multiple 
of $2\pi$. We define this core as a
giant vortex. Similar structures have been observed in experiments on 
rapidly rotating
superflow in BECs.~\cite{exper} It should be noted that some 
authors reserve the term giant vortex for multiply quantized 
vortices,~\cite{kanda,stopa} such as those found in the Laughlin 
states.~\cite{laughlin} Our intention is, however, to analyze a 
phenomenon which is analogous to the formation of giant
vortices in BECs.

In this paper we report theoretical prediction that giant-vortex
structures can be found in rotating quasi-2D 
fermion systems. These states are analogous 
to those found in BECs, and they emerge when anharmonicity
(flatness) is added in the otherwise parabolic confining potential.
Unlike in the bosonic case, the tendency of giant
vortex formation is limited to systems with a low number of electrons
($N<20$). As the second difference, the cores of giant
vortices in QDs do not merge completely. We show this effect
using conditional wave functions to probe the internal 
structure of the many-electron state. We also suggest how the 
formation of giant vortices could be observed in experiments. 

\section{Model}

Our model system is a QD with electrons being restricted in
quasi-2D plane and interacting via their mutual Coulomb
repulsion. The Hamiltonian is then
\begin{equation} H=\sum^N_{i=1}\left[\frac{(-i\hbar
\nabla_i+e {\bf A} )^2}{2 m^*} +V_{{\rm c}}(r_i)\right] + \frac{e^2}{4\pi
\epsilon} \sum_{i<j} \frac{1}{r_{ij}} \ , \label{hamiltonian}
\end{equation}
where $N$ is the number of electrons, ${V_{\rm c}}$ is
the external confining potential on the plane, and
${\bf A}$ is the vector potential of the homogeneous magnetic field
oriented perpendicular to the QD plane.
We imitate the conditions of a real semiconductor 
heterostructure by using the effective material parameters 
for GaAs, i.e., the effective mass $m^*=0.067\,m_e$ and 
the dielectric constant $\epsilon=12.4\,\epsilon_0$.

We solve the many-particle Schr\"odinger equation for the Hamiltonian 
(\ref{hamiltonian}) using two numerical approaches, namely the 
configuration interaction (CI) method which numerically
diagonalizes the Hamiltonian, and the
spin-density-functional theory (SDFT) 
in a real-space formulation.~\cite{mika}
We point out that our previous studies~\cite{henri1,esarecta} show that  
SDFT provides a proper description of the high-field solutions 
even if the total energies given by the computationally more demanding  
current-spin-density-functional theory are slightly closer 
to the quantum Monte Carlo results.~\cite{csdft} 

According to theoretical
studies of rotating BECs, giant-vortex structures may emerge in
bosonic systems when the confining potential ${V_{\rm c}}$ has a flat,
non-parabolic shape.~\cite{lundh,kasamatsu,jackson1}
Therefore, within the CI method we apply a confinement
following the notation of Jackson and co-workers:~\cite{jackson1} 
\begin{equation} 
V_{\rm c}^{\rm quartic}(r)=\frac{1}{2}m^* \omega_0^2 r^2\left
[1+\lambda\left(\frac{r}{a_0}\right)^2\right].
\label{quartic}
\end{equation}
Here $\hbar\omega_0$ is the confinement strength fixed to
$5$ meV unless stated otherwise, 
$a_0=\sqrt{\hbar/m^*\omega_0}\approx 15.1$ nm is the oscillator 
length, and $\lambda$ is a positive dimensionless constant determining  
the strength of the quartic contribution, i.e., the 
flatness of the confinement. 
Within the SDFT approach we
apply also a circular hard-wall potential well which can be
considered as the extreme limit of the flatness of the
potential. It is defined simply as 
\begin{equation}
V_{\rm c}^{\rm well}(r)=\left\{ \begin{array}{ll}
0, & r\leq R \\
\infty, & r>R,
\end{array} \right.
\label{well}
\end{equation}
where $R$ is the radius of the well. 
It should be noted that 
as a reasonable approximation, the confining potential in 
semiconductor QDs is usually assumed to be parabolic, although
detailed modeling of realistic QD systems has pointed out the 
importance of anharmonicity in the potential.~\cite{matagne} 
In addition, steep edges have been found to explain observed 
features in far-infrared absorption spectra.~\cite{ullrich}
As we shall see below, even small deviations from the harmonicity 
may lead to emergence of new types of ground states.

\section{Phase diagram}

\begin{figure} 
\includegraphics[width=0.9\columnwidth]{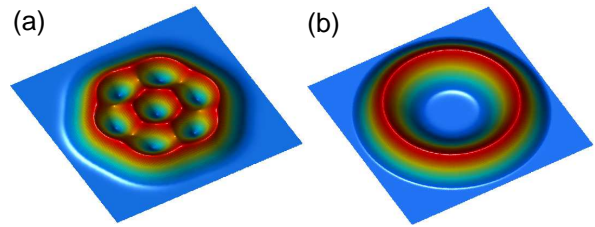} 
\caption{(Color online)
Electron densities (SDFT) of
$N=10$ quantum dots containing seven flux 
quanta in both but having different
distributions of vorticity. The state in (a) is
calculated using parabolic external potential
($\hbar \omega=4\;{\rm meV}$, $B=13$ T) and shows a cluster
of seven vortex holes. The state in (b) is calculated in hard-wall
potential well ($R=100\;{\rm nm}$, $B=25$ T)
and shows a giant core (eye) of a vortex comprising seven flux quanta close
to the center of the rotation.}
\label{giant}
\end{figure}

First we visualize the fundamental difference between 
the high-field solutions corresponding to a vortex 
cluster~\cite{henri1} and a giant vortex in a QD.
Figure~\ref{giant} 
shows the SDFT results for electron density of a 
vortex-cluster state (a) and a giant-vortex state (b) 
in a 10-electron QD defined by a parabolic ($\lambda=0$) and
circular hard-wall confining potentials, respectively.
In the parabolic case the vorticity is spread throughout the 
system which gives rise to a cluster of seven hole-like 
quasi-particles.~\cite{henri1,manninen}
If the confinement is set flat, however, we find 
a large core in electron and current densities at the
center. In this case all phase singularities  are absorbed 
into the core region to form a single seven-fold giant vortex.
The qualitative origin of the phenomenon is in the fact
that increasing anharmonicity (flatness) in the
external confinement leads to a relative increase in the energetic
advantage of the outer orbitals compared to the inner ones. In
consequence, there is rearrangement in the electron occupations 
such that the inner orbitals are left empty (or they have a 
negligible weight). Correspondingly, the vortices (holes in the 
electron density) are accumulated at the center.

\begin{figure} 
\includegraphics[width=8cm]{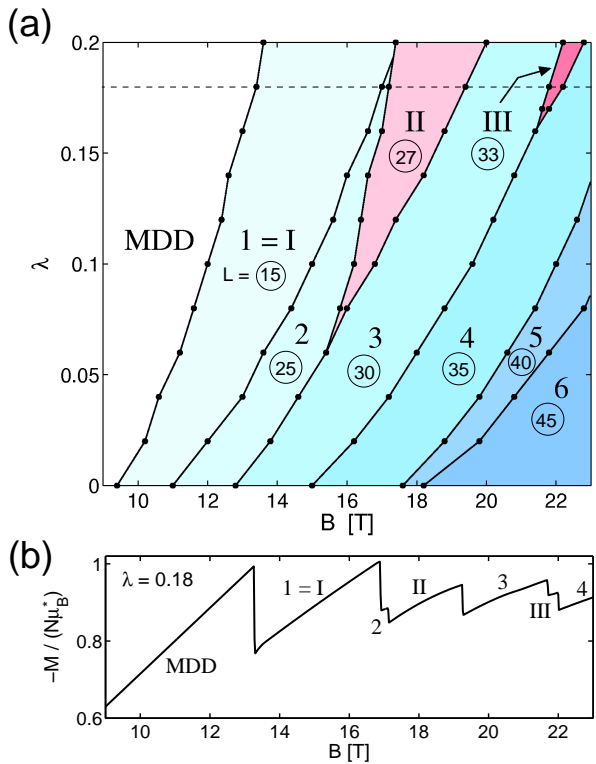} 
\caption{(Color online) (a) Phase diagram of the vortex  
solutions in a six-electron quantum dot defined by a 
parabolic-plus-quartic
confining potential [Eq. (\ref{quartic})].
The results have been calculated with the CI method.  
The Roman and Arabic numerals mark 
the number of multiple and clustered vortices in 
the ground states, respectively. The total angular
momenta $L$ are marked by circled numbers.
(b) Magnetization of a quantum dot with $\lambda=0.18$ 
corresponding to the dashed line in (a). 
} 
\label{phase} 
\end{figure}

To gain more physical insight into giant vortices,
we present in the following a detailed analysis of the vortex states in a
six-electron QD. In a parabolic potential the calculations 
have shown that the ground states in the fractional
quantum Hall regime (filling factor $\nu<1$) occur only
at certain magic values for the angular 
momentum $L$ (Ref.~\onlinecite{maksym}).
For $N=6$ the $\nu=1$ state corresponds to
$L=15$ maximum-density droplet (MDD) and the subsequent magic
angular momenta are $L=21,\, 25,\, 30,\, 35,$...
These states correspond to ground states
with an increasing number of off-electron vortices forming 
clusters inside the electron droplet.~\cite{henri1}
When the confining potential is made flatter by including
the quartic contribution of Eq.~(\ref{quartic}),
new ground states emerge corresponding to giant 
vortices.
The CI calculations show that when $\lambda>0.06$ a 
double vortex core appears between $L=25$ and $L=30$ states 
and it has an angular
momentum $L=27$. Triple vortex core emerges for
$\lambda>0.16$ and it has $L=33$.
This is illustrated in Fig.~\ref{phase}(a) 
which shows the phase diagram
of the CI solutions for $0<\lambda<0.2$.
It should be noted that the minimum value
of the quartic multiplier $\lambda$ required
for giant vortices is relatively small 
($\lambda \approx 0.06$), so that 
the results may apply to a large class of real 
quantum dot devices. 
Fig.~\ref{phase}(a) shows also
that the giant-vortex states vanish rapidly at high fields  
unless $\lambda$ is considerably increased. This is due to the  
increasing magnetic confinement which has a parabolic form.  

Fig.~\ref{phase}(b) shows the magnetization 
$M=-\partial E_{\rm tot}/\partial B$ of a QD with  
$\lambda=0.18$ associated with the dashed line in Fig.~\ref{phase}(a). 
The steps in $M$ correspond to changes in the 
many-electron ground state {\it either} due to an increase in 
the vortex number \cite{henri4} {\it or} due to a rearrangement 
of the vortices (and electrons) in the system.
The oscillative behavior of the magnetization might be observable 
in, e.g., magnetization measurements for large 
ensembles of QDs. Such experiments can been done using 
sensitive micromechanical magnetometers.~\cite{markkumusta} 
Our data suggest, however, that the ground-state transitions 
associated with the rearrangement of vortices and electrons 
usually cause relatively small changes in the magnetization.
Hence, the observation of such transitions would require a high 
degree of accuracy in the experiments.

\section{Internal structure of giant vortices}

\begin{figure} 
\includegraphics[width=7.5cm]{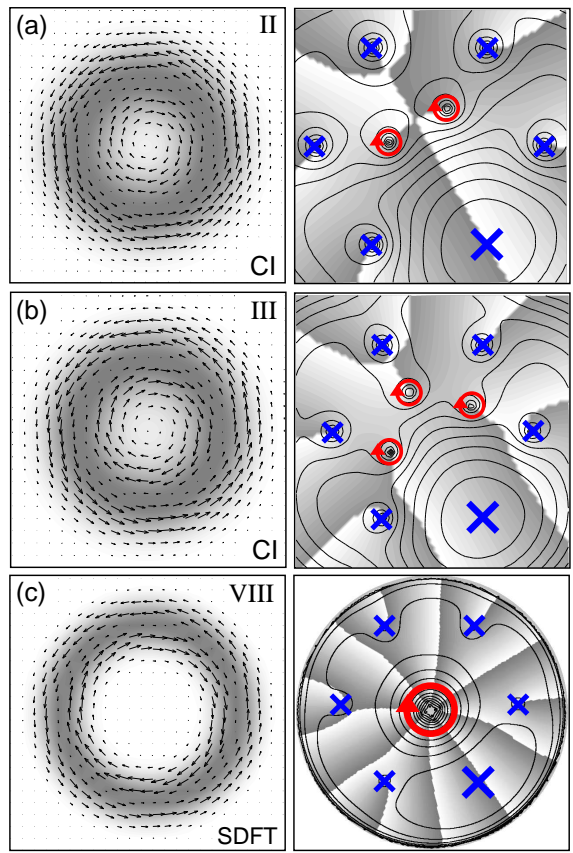} 
\caption{(Color online) 
Left panel: 
Total electron densities (grayscale)
and current densities (arrows) 
of double-vortex (II),
triple-vortex (III), and eight-fold giant-vortex (VIII)
states in a six-electron quantum dot. The states II and
III have been calculated with the CI in a
quadratic-plus-quartic potential as in Fig.~\ref{phase}.
The state VIII has been calculated with the SDFT in
a circular hard-wall well of radius $R=70$ nm.
Right panel:
Corresponding conditional densities (contours) and 
the phases (grayscale) of the conditional wave function.
The phase of the SDFT result (VIII) is calculated
from the single-determinant auxiliary wave function constructed 
from the Kohn-Sham states. At the lines where  
the shadowing changes from darkest gray to white the 
phase jumps from $-\pi$ to $\pi$. 
The crosses (blue) and circles (red) mark the most probable  
electron positions and vortex positions, respectively. 
The Pauli vortices at the electron positions
are not marked by circles for clarity. 
The probing electron has its maximum probability downright.
} 
\label{cond} 
\end{figure}

In order to visualize the internal structure of the giant-vortex
states, we compute the conditional wave 
functions~\cite{henri1} from the numerically exact
many-body ground states. Within the SDFT instead, we use the auxiliary
single-determinant wave function constructed from Kohn-Sham
states.~\cite{henriphysica,ellipse}
Figure~\ref{cond} 
shows the total electron and current densities (left panel)
and the conditional wave functions (right panel)
of the double (a) and triple (b) vortex states at $L=27$ and $L=33$ 
(CI) and a giant eight-fold vortex state (c) in a circular well with $L=63$ 
(SDFT). Since the system is totally polarized there is a Pauli vortex on
each electron position as mandated by the exclusion principle.
The exchange-hole around each electron can be associated with these vortices.
In addition, there are off-electron vortices close to the
center of rotation. They lead to holes in the electron density
with rotational current around them. The size of the vortex hole at
the center of the QD increases with the number of vortices as shown in
the left panel of Fig.~\ref{cond}.

Despite the formation of single cores in the electron and current 
densities, the conditional wave functions [right parts of 
Figs.~\ref{cond}(a) and (b)] show finite separation 
between the vortices at the center. 
The separation is due to quantum fluctuations appearing as 
effectively repulsive vortex-vortex interactions. 
This effect can also be interpreted as zero-point motion of
vortices,~\cite{ellipse} which leads to non-zero electron density at
at the center of the giant vortex. In the bosonic case the vortex-vortex 
interaction energy as a function of vortex separation
has been approximated to be of logarithmic form which prevents
complete merging of the vortex cores.~\cite{kasamatsu} 

In the SDFT result for the eight-fold giant vortex 
[Fig.~\ref{cond}(c)], the density is zero at the center
and very small up to the QD radius where the electrons are 
strongly localized. The vortices at the center are then packed
close to each other but still, however, showing a small-scale
structure. The stronger vortex localization in the SDFT 
can be understood from the fact that the SDFT generally can 
incorporate less correlation effects than exact many-body methods. 

Qualitatively similar internal structures have been found for giant 
vortices in rotating BECs.~\cite{kasamatsu}
In that system, however, a local potential maximum at 
the center attracts the phase singularities into a single 
point in space so that a {\em multiply quantized} vortex is 
formed.~\cite{jackson1,fischer}
This is not the case in QDs, where the confining potentials 
increase monotonically as a function of the radius of the system.
In QDs a pinning potential, expectedly leading to perfect multiple
phase quantization like in bosonic systems, could be realizable 
in form of a negatively charged impurity at the center of the sample.


\section{Dependence on the electron number}

\begin{figure}
\includegraphics[width=7cm]{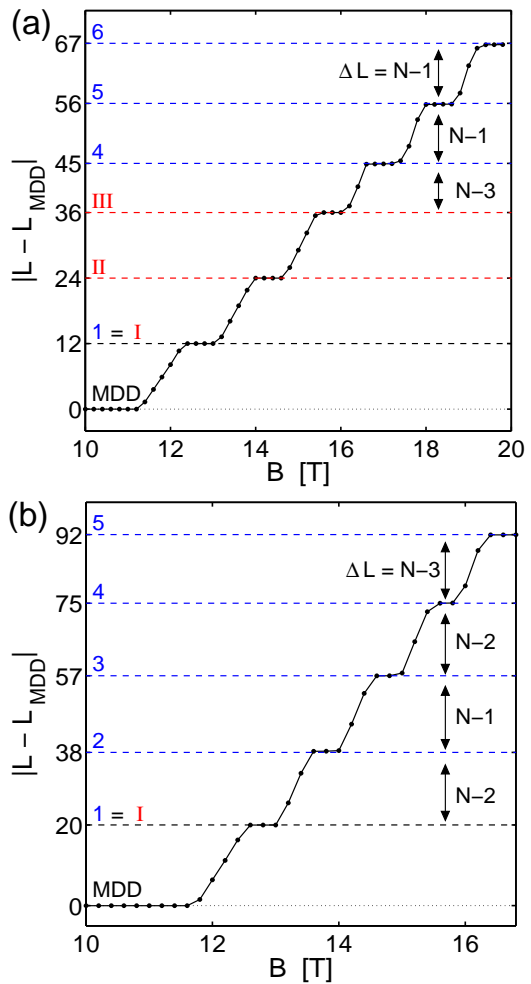}
\caption{(Color online) SDFT results for the total angular momenta 
of $N=12$ (a) and $N=20$
(b) hard-wall quantum dots.
Red and blue dashed lines mark the
$L$ plateaus for multiple vortices (Roman numerals) and vortex
clusters (Arabic numerals), respectively.
The black dashed line corresponds to the single-vortex solution.}
\label{many}
\end{figure}

When $N$ is increased the maximum number of flux quanta that can 
be packed to the center of a giant vortex decreases rapidly. 
According to our SDFT calculations for hard-wall QDs 
[Eq.~(\ref{well})] of realistic radii ($R=50\ldots 150$ nm), 
giant vortices consisting of more than ten flux quanta can
be found when the electron number is increased up to $N=10$. 
When $N=12$, however, only three flux quanta can be
packed at the center. For $N>18$ giant vortices 
disappear completely for realistic well 
sizes, and they are replaced by vortex clusters inside the
electron droplet. We visualize this effect in Fig.~\ref{many},
which shows the total angular momenta of the SDFT solutions
for $N=12$ (a) and $N=20$ (b) hard-wall QDs 
as a function of the magnetic field. 
Red and blue dashed lines mark the
$L$ plateaus for multiple vortices (Roman numerals) and vortex
clusters (Arabic numerals), respectively, and the black dashed
line corresponds to the single-vortex solution. The results 
between the plateaus are a consequence of mixing of the consecutive 
exact ground states.~\cite{henri1} 
In the range of the MDD and the highest 
multiple vortex solution, the differences in $L$ between 
the plateaus are $\Delta L=N$. This regularity
follows from the simple increase in the hole size discussed above.
When the vortices 
become clustered instead, we find smaller steps in $L$ resulting
from the complex rearrangements of vortex clusters in the QD.

The disappearance of giant vortices as a function of $N$
is related to the formation of shell structure in the QD, 
i.e., it becomes energetically favorable for
the electrons to occupy the inner orbitals of the system.
This behavior is in striking contrast with what is
found in bosonic systems where giant vortices can be found in
BECs with a very large number of particles. 
On the other hand, our SDFT results are consistent with the 
exact diagonalization results which show that
even single vortices tend not to localize at the center of a 
parabolic QD when $N>12$ (Ref.~\onlinecite{AriLowTempPhys}).

\section{Summary}

To summarize, on the grounds of our numerical
analysis we predict that giant vortices emerge in
quantum dots in high magnetic fields, or more generally, in rapidly
rotating electron droplets.
These ground states are analogous to those found in rotating
Bose-Einstein condensates. In quantum dots only a slight
anharmonicity (flatness) in the external confining potential
is required for the formation of giant vortices.
By using conditional wave functions we can analyze the internal 
structure of the vortex cores. Formation of giant vortices leave
measurable signatures in the ground state electron and current densities
as well as in the magnetization of quantum dots.
These signatures can, in principle, be detected in state-of-the-art
magnetization or electron transport measurements or via direct imaging
of the charge density in quantum dots.
As in the bosonic case, the conditions for the emergence of giant-vortex
states are determined by the interplay between the interactions and
the form of the confining potential. We find that in quantum dots the 
giant-vortex solutions are generally limited to systems consisting of 
less than 20 electrons. However, our numerical analysis has been done assuming
the conventional electron-electron interaction of a Coulombic form. 
Other types of interparticle potentials may lead to quantitatively
different results, but we leave this topic for future research.

\begin{acknowledgments} 
We thank Matti Manninen and Georgios Kavoulakis for helpful discussions.  
This research has been supported by the Academy of Finland through 
its Centers of Excellence Program (2000-2005), the Swedish Research Council 
and the Swedish Foundation for Strategic Research.
E.R. acknowledges support from the NANOQUANTA NOE and from the Finnish
Academy of Science and Letters,
Vilho, Yrj{\"o} and Kalle V{\"a}is{\"a}l{\"a} Foundation.

\end{acknowledgments}

\end{document}